\documentclass[preprint,12pt]{elsarticle}

\usepackage{amssymb}
\usepackage{url}

\usepackage[nodots,nocompress]{numcompress}

\journal{Nuclear Instruments and Methods in Physics Research A }

\begin{document}

\begin{frontmatter}



\title{PoGOLino: a scintillator-based balloon-borne neutron detector}

\author[label1,label2]{Merlin Kole\corref{cor1}}
\author[label1,label2]{Maxime Chauvin}
\author[label3]{Yasushi Fukazawa}
\author[label4]{Kentaro Fukuda}
\author[label4]{Sumito Ishizu}
\author[label1,label2]{Miranda Jackson}
\author[label5]{Tune Kamae}
\author[label4]{Noriaki Kawaguchi}
\author[label3]{Takafumi Kawano}
\author[label1,label2]{M\'ozsi Kiss}
\author[label1,label2]{Elena Moretti}
\author[label1,label2]{Mark Pearce}
\author[label1,label2]{Stefan Rydstr\"om}
\author[label3]{Hiromitsu Takahashi}
\author[label6]{Takayuki Yanagida}

\cortext[cor1]{Corresponding author. Tel.: +46 85 537 8186 ; fax: +46 85 537 8216. E-mail address: merlin@particle.kth.se}

\address[label1]{KTH Royal Institute of Technology, Department of Physics, 106 91 Stockholm, Sweden}
\address[label2]{The Oskar Klein Centre for Cosmoparticle Physics, AlbaNova University Centre, 106 91 Stockholm, Sweden}
\address[label3]{Department of Physical Science, Hiroshima University, Hiroshima 739-8526, Japan}
\address[label4]{Tokuyama Corporation, Shunan, Yamaguchi, Japan}
\address[label5]{University of Tokyo, Deptartment of Physics, 113-0033 Tokyo, Japan}
\address[label6]{Kyushu Institute of Technology, Kitakyushu, Fukuoka, Japan}

\begin{abstract}

PoGOLino is a balloon-borne scintillator-based experiment developed to study the largely unexplored high altitude neutron environment at high geomagnetic latitudes. The instrument comprises two detectors that make use of LiCAF, a novel neutron sensitive scintillator, sandwiched by BGO crystals for background reduction. The experiment was launched on March 20th 2013 from the Esrange Space Centre, Northern Sweden (geomagnetic latitude of $65^\circ$), for a three hour flight during which the instrument took data up to an altitude of 30.9 km. The detector design and ground calibration results are presented together with the measurement results from the balloon flight. 

\end{abstract}


\end{frontmatter}


\section{Introduction}

A detailed understanding of the atmospheric neutron environment is necessary for radiation protection studies linked to air travel and space tourism, as well as the phenomena of single event upsets in avionics \cite{SEU}. Furthermore, for Earth-orbiting and balloon-borne experiments making use of low Z detection materials, high energy neutrons may produce an irreducible background when scattering in the detector. Data on of the neutron environment at high latitude balloon altitudes is however lacking. An example of an experiment subject to an irreducible neutron induced background is the PoGOLite instrument of which PoGOLino is a spin-off \cite{PoGOLite}. PoGOLite is a balloon-borne X-ray polarimeter which makes use of plastic scintillators to detect both the scattering and photoabsorption location of incoming X-rays. Plastic scintillators have a high Compton scattering cross section for hard X-rays, but also for elastic neutron scattering. At the float altitude of PoGOLite, 40 km, the background induced by neutrons is larger than the expected signal despite the use of a 300 kg passive polyethylene shield used to reduce the incoming high energy neutron flux. Due to this irreducible background and the long duration circumpolar flight path of the instrument, a detailed knowledge of the neutron environment at high altitudes and its dependencies on location and solar activity is of great importance. 

Neutrons are produced in the Earth's atmosphere in air showers resulting from interactions of cosmic rays with atmospheric nuclei. Within the shower neutrons are produced either by weak interaction processes, in internal cascades in the nuclei or by the evaporation of these nuclei after the initial interaction. The production energy of neutrons from these processes range from as low as $\sim0.1\,\mathrm{MeV}$, for evaporation, to a maximum of 10's of GeV for weak interaction processes. After production, the neutrons are moderated to sub-eV energies by scattering off atmospheric nuclei. The different production processes and the atmospheric moderation result in a complex relationship between the neutron energy spectra, directional dependence and altitude. The neutron environment furthermore depends on the incoming cosmic ray spectrum which in turn varies with latitude and solar activity, further adding to the complexity of the atmospheric neutron environment. 

As a result of the large number of dependencies on the neutron environment, accurate predictions of it rely on Monte Carlo simulations which are validated using experimental data. For low latitudes such high altitude data exists as a result of, for example, several measurements performed during balloon missions launched from Palestine, Texas \cite{E2} \cite{E3}. Experimental data at high altitudes is however lacking for high latitudes where, as a result of the Earth's magnetic field, the neutron flux is highest and most sensitive to solar activity. Interest in accurate predictions for these altitudes is further increased as a result of the recent popularity of balloon flights from high latitude locations such as the McMurdo Station in Antarctica and the Esrange Space Centre in Northern Sweden. Acquiring more data of the neutron environment at these locations is therefore necessary in order to validate and develop Monte Carlo based atmospheric neutron models. 

The increasing price of $\mathrm{^3He}$, the traditional material used for neutron detection, has prompted the development of novel neutron detectors. Among these, the recently developed Lithium Calcium Aluminium Fluoride (LiCAF) scintillator crystal allows compact, mechanically simple, lightweight and efficient detector systems to be developed. LiCAF scintillator crystals can be combined with high Z scintillators in a Phoswich (phosphorous sandwich) configuration, a combination of several different scintillators read out using a single photomultiplier tube (PMT), to reduce instrumental background. The high detection efficiency per unit volume allows for a small detector whereas the scintillation crystal-based detection mechanism allows for a mechanically simple instrument with respect to, for example, gas-based detectors. This combination of features makes LiCAF ideal for balloon-borne neutron measurements. 

The PoGOLino instrument contains two neutron detectors, one of which is embedded in a moderating material, resulting in a high efficiency for high energy neutrons whereas the second detector remains unshielded resulting in a high efficiency at low energies, thereby giving insight into the spectral shape of the incoming spectrum. By keeping the instrument lightweight and designing it for autonomous operation, it is optimised for piggyback flights on other balloon experiments.

PoGOLino was launched on March 20th 2013 from the Esrange Space Centre in Northern Sweden. Data was collected both during the $\sim2$ hour ascent phase and during the additional 1 hour spent at the float altitude of 30.9 km. 

In the following section a detailed description of the instrument will be provided, followed by results of the on-ground calibration measurements. The 2013 flight is then described and the flight measurement results are compared to simulated data.

\section{Instrument design}

\begin{figure}[!b]
  \begin{center}
    \includegraphics[width=0.58\textwidth]{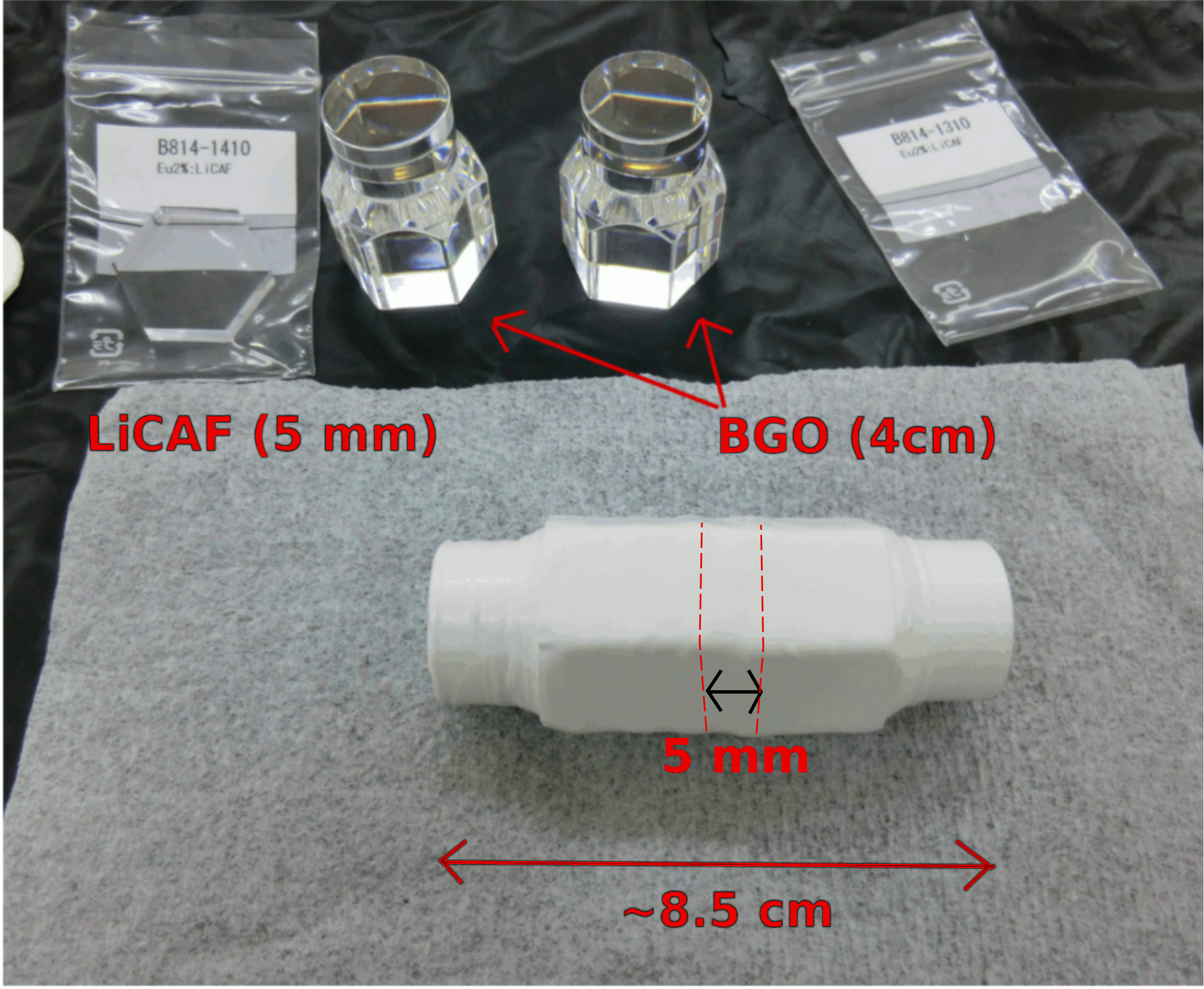}
  \end{center}
  \caption{An assembled PDC in front of the three individual crystals.}
  \label{PDC}
\end{figure}

LiCAF was chosen as the scintillator material for the PoGOLino experiment to maintain a low mass while ensuring a high neutron detection efficiency. LiCAF is a novel, low Z, inorganic scintillator material with a high $^6\mathrm{Li}$ content developed at Tokuyama corporation, Japan. Neutron detection proceeds through the reaction $^6\mathrm{Li} + \mathrm{n} \rightarrow\,^4\mathrm{He}\,(2.73\,\mathrm{MeV}) +\,^3\mathrm{T}\,(2.05\,\mathrm{MeV})$. The cross section for this process for thermal neutrons is 940 barn. The low Z value of the material, which results in a relatively low cross section for photon interactions, and the monoenergetic signal from neutron capture makes LiCAF suitable for performing pulse height based photon/neutron discrimination. Both a cerium and a europium doped version of LiCAF exists. Cerium doped LiCAF has a decay time of 40 ns and a light yield of $3.5\times 10^3$ photons/neutron capture \cite{Ce}. Europium doped LiCAF has a longer decay time, 1600 ns, but has the advantage of a high light yield of $3\times 10^4$ photons/neutron capture \cite{Eu}. Due to the relatively low counting rates expected at float altitudes, on the order of 10 Hz, a short decay time is not required for PoGOLino. Europium doped LiCAF, which has the advantage of a higher light yield was therefore chosen as a detection material. The emission wavelength from Eu:LiCAF is 370 nm making it compatible with Bismuth Germanium Oxide (BGO) scintillators in a Phoswich configuration. In the Phoswich Detector Cells (PDCs) used in PoGOLino, a hexagonal LiCAF scintillator with a thickness of $5\,\mathrm{mm}$ and a side length of $16\,\mathrm{mm}$ is used. The high detection efficiency per unit volume of LiCAF  ensures an expected total number of neutron detections of order $10^3$ for a 300 s measurement for this size of scintillator. Assuming a balloon ascent velocity of $1\,\mathrm{km}/300\mathrm{s}$ this will enable PoGOLino to perform neutron flux measurements with a statistical error on the order of a few percent within a 1 km ascent. To reduce instrumental background from photons and charged particles the LiCAF scintillator is placed between two $40\,\mathrm{mm}$ thick BGO crystals which form an anticoincidence system. The BGO crystals used for PoGOLino were produced by the Nikolaev Institute of Inorganic Chemistry. BGO was chosen for its high Z value which results in a high photoabsorption cross section. Since the decay time of BGO emission is 300 ns, pulse shape discrimination can be used to distinguish between the LiCAF events and BGO events. The assembly of 3 scintillator crystals is read out by a PMT of the type Hamamatsu Photonics R7899EGKNP originally developed for the PoGOLite mission \cite{Tune}. Figure \ref{PDC} shows an assembled PDC together with the individual components. PoGOLino contains two such PDCs. One is shielded by a polyethylene cylinder with a radius of 8.5 cm while the other remains unshielded. The unshielded detector is sensitive to thermal neutrons whereas, due to neutron moderation by the polyethylene, the shielded detector is sensitive to neutrons over a wider energy range, thereby making PoGOLino sensitive in two different energy regimes. Aside from these two PDCs, the PoGOLino instrument has a third PDC in which the LiCAF crystal is replaced by a plastic EJ-204 scintillator, produced by Eljen Technology with the same geometry as the LiCAF crystals. The data from this PDC will be used for the study of neutron interactions in EJ-204. Such interactions are expected to form the main background for the polarisation measurements of the PoGOLite experiment \cite{Merlin_thesis}. 

\begin{figure}[!b]
  \begin{center}
    \includegraphics[width=1.0\textwidth]{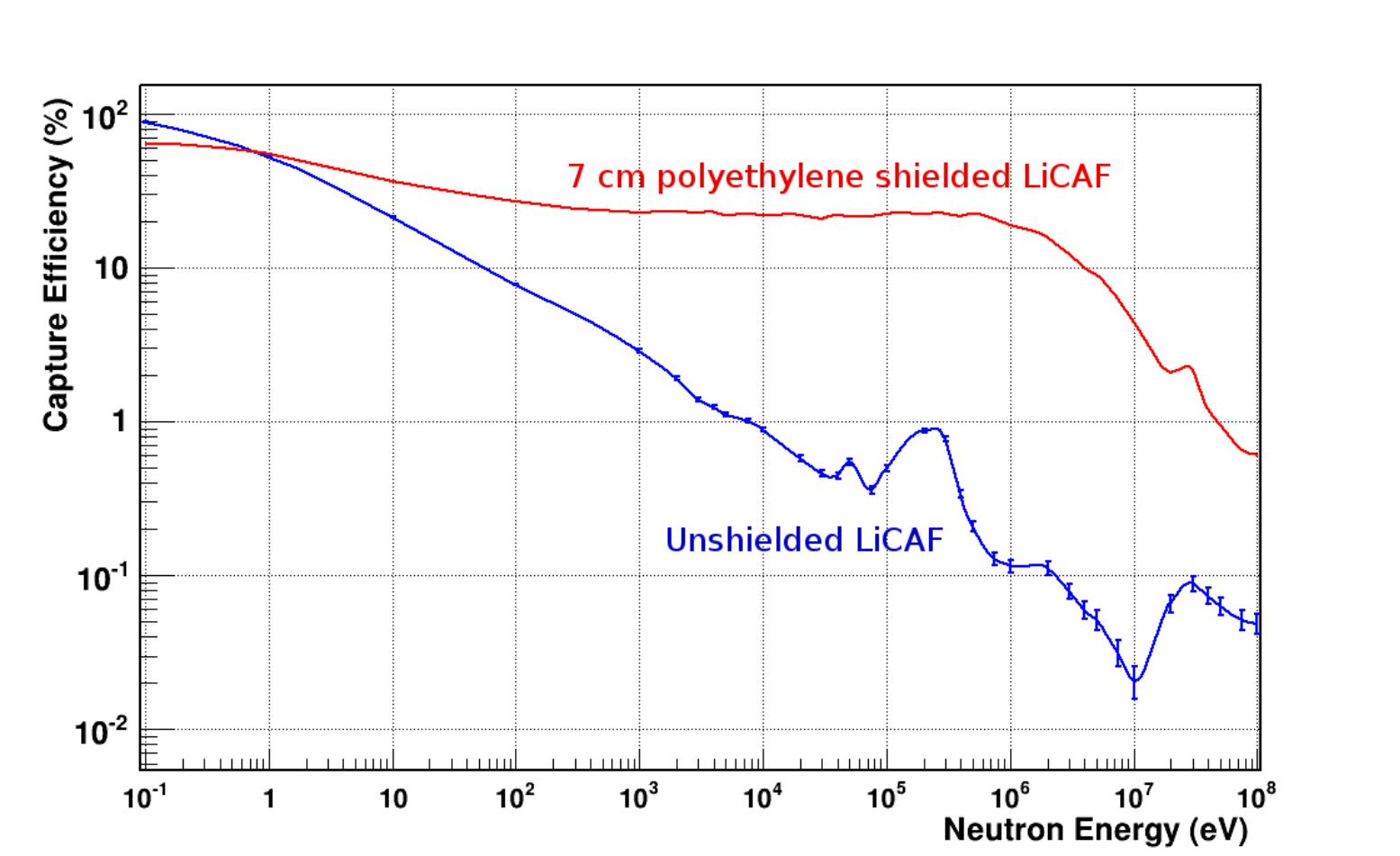}
  \end{center}
  \caption{The neutron capture efficiency as a function of energy for the shielded (red) and unshielded (blue) LiCAF-BGO PDC as simulated using Geant4.}
  \label{cap_eff}
\end{figure}

The energy dependence of the neutron capture efficiency of the two LiCAF-BGO PDCs was simulated using Geant4 \cite{Geant4}, a C++ based software toolkit developed to simulate particle interactions in matter. The version of Geant4 used for this work is 4.9.6. This version was used with the QGSP\_BIC\_HP physics list together with the neutron high precision thermal scattering data libraries. Results are presented in figure \ref{cap_eff}. Due to the limited moderation of the LiCAF material itself the neutron capture cross section as a function of energy of the unshielded detector, shown in blue, is similar to that of pure $^6\mathrm{Li}$. The efficiency therefore peaks below $1\,\mathrm{eV}$: the thermal energy range. For the other PDC the polyethylene serves to moderate high energy neutrons towards lower energies while reducing the incoming thermal neutron flux. As a result, the shielded detector has a capture cross section, shown in red, which is almost constant with energy from the thermal range up to MeV energies, after which the cross section quickly drops off. By measuring in two different energy regimes, PoGOLino is be able to measure the energy-integrated atmospheric neutron flux and give information on the shape of the energy spectrum. 

The three PMTs are read out using a Flash ADC board (FADC) developed for the PoGOLite project with a SpaceWire interface \cite{HT}. PMT waveforms are sampled at a rate of $37.5\,\mathrm{MHz}$. Whereas this sampling rate is necessary for waveform discrimination between signals from the plastic and the BGO scintillators, waveform discrimination between BGO and LiCAF signals can be achieved with a lower sampling rate. In order to save memory for the LiCAF-BGO channels only every third clock cycle therefore needs to be stored for these two channels, resulting in an effective sampling rate of $12.5\,\mathrm{MHz}$.  A trigger is issued when the Analog-to-Digital Converter (ADC) level exceeds a predefined threshold level. For each trigger the ADC levels measured during the previous 10 sampled clock cycles and those of the following 40 sampled clock cycles are stored. In order to reduce deadtime, a simple online pulse shape discrimination is performed by the Field Programmable Gate Array (FPGA). This discrimination, between BGO and LiCAF events, is based on the ratio between the ADC levels measured at the 4th and 19th stored clock-cycle after a trigger is issued. Here the 4th and 19th stored clock-cycle correspond to the expected peak position of the waveforms of BGO and LiCAF respectively. In order to further improve the signal-to-background ratio, additional pulse shape discrimination can be applied offline. To reduce deadtime during flight, the majority of the data was taken while applying online vetoes. In order to verify the veto performance, however, some data sets were acquired without the application of vetoes. Apart from the waveforms the ADC values measured at either the 4th or the 19th clock cycle is stored additionally after a trigger is issued independently from the waveform data, thereby making it independent on the deadtime induced by the storage of waveforms. Whether the 4th or the 19th ADC value is stored depends again on the ratio of these two values, the used calculation and selection criteria are equal to that used for the online veto application. The final result is a set of two histograms, one with all BGO induced triggers and one with all LiCAF induced triggers. These histograms can be used to verify the deadtime corrected counting rate as measured using the waveform data. 

\begin{figure}[!b]
  \begin{center}
    \includegraphics[width=0.9\textwidth]{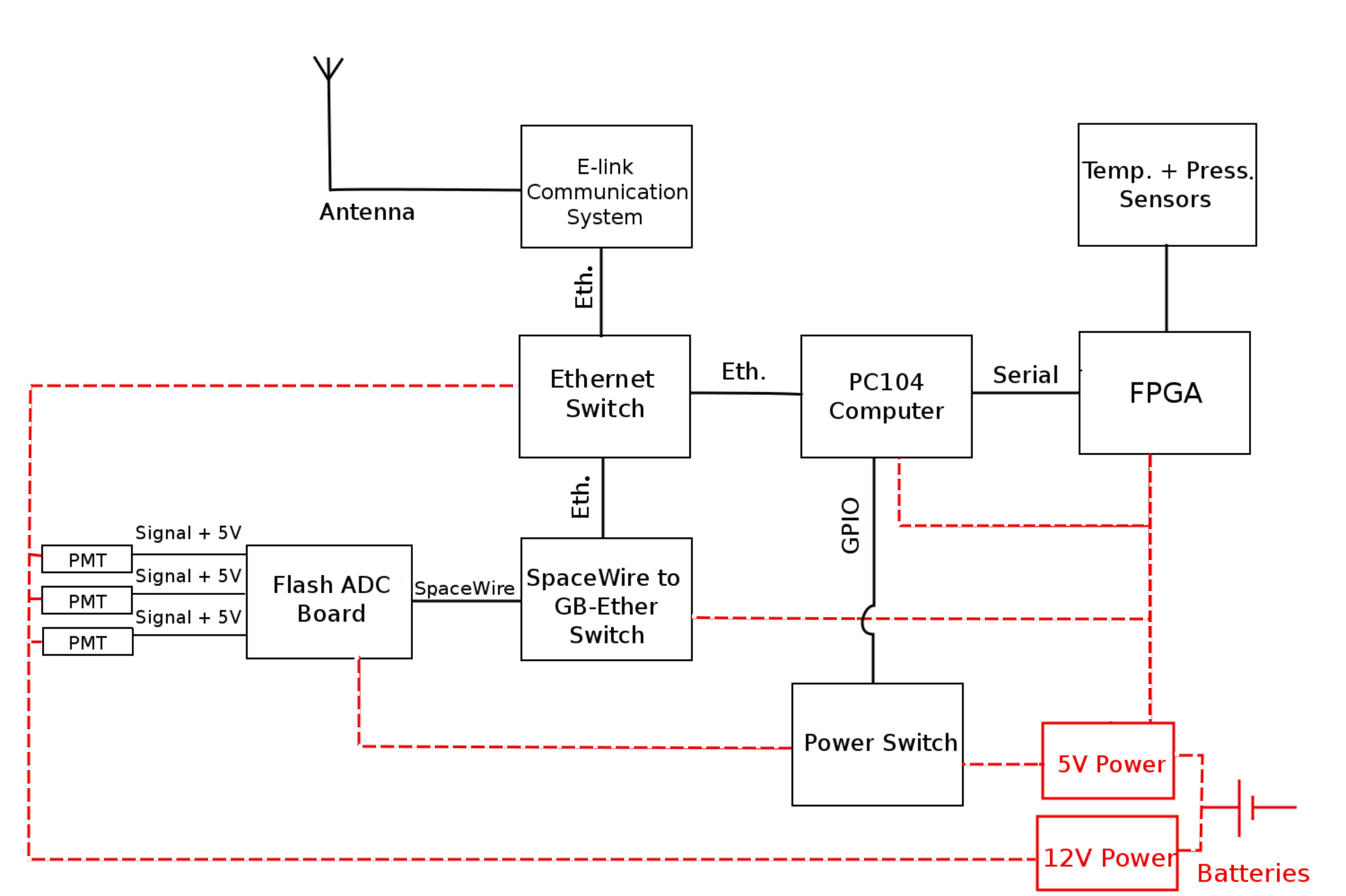}
  \end{center}
  \caption{A schematic block diagram of the PoGOLino instrument with the different signal connections shown as full black lines and the 12V and 5V power supplies in red dashed lines.}
  \label{diagram}
\end{figure}

\begin{figure}[!b]
  \begin{center}
    \includegraphics[width=0.58\textwidth]{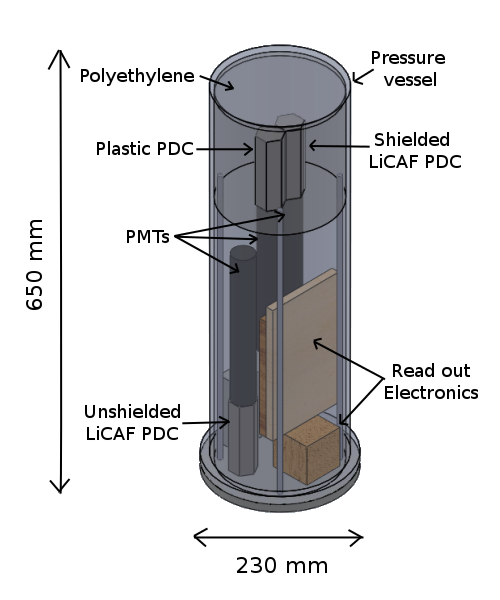}
  \end{center}
  \caption{A schematic overview of the PoGOLino instrument.}
  \label{full_ins}
\end{figure}

The FADC board is controlled and read out through a SpaceWire to Gigabit Ethernet Bridge by a PC104 computer. During flight, connection to the PC104 computer was possible through an Ethernet-over-radio system (E-link) provided by the Esrange Space Centre \cite{Esrange}. To mitigate a possible loss of communication the PC104 computer was programmed to control the FADC board such that it performs data taking autonomously. The functionality of the FADC board is checked every 10 minutes. In case of data acquisition problems, the FADC board is automatically power-cycled using a switch controlled by the General-Purpose Input/Output (GPIO) of the PC104 computer, as can be seen in the system diagram in figure \ref{diagram}. During standard operation, the waveforms and trigger rates from the FADC are written to a solid state disk connected to the PC104 computer. The PC104 computer is also used to read out three temperature sensors and two pressure sensors through a Field Programmable Gate Array (FPGA). Two of the temperature sensors are placed on different PDCs while the third is placed among the read out electronics. The pressure sensors are placed on opposite sides of the instrument in order to provide additional information on a potential temperature gradient in the instrument. Data on the temperature of the instrument during the entire flight is essential due to the temperature dependence of the BGO light yield and decay time and the temperature dependence of the scattering cross sections for low energy neutrons. Mechanical support and temperature stability during flight is provided by placing all equipment inside a $670\,\mathrm{mm}$ high cylindrical aluminium pressure vessel with a radius of $165\,\mathrm{mm}$ and wall thickness of $3\,\mathrm{mm}$, as shown in figure \ref{full_ins}. An autonomous heating system in the pressure vessel provides $\sim30\mathrm{W}$ of heating when temperatures inside the instrument drop below $0\,^\circ\mathrm{C}$. 

An operational time of 6 hours is provided by 8 batteries, type SAFT LSH20, which power the instrument. The batteries provide power both to a 12 V and a 5 V circuit through DC/DC converters. The internal DC/DC converters in the PMTs are supplied with 12 V directly from the battery system and with an additional control voltage between 0 and 5 V provided through and controlled by the FADC board. The mass of PoGOLino is 13 kg. The system diagram of the instrument can be seen in figure \ref{diagram}, a schematic overview of the instrument is shown in figure \ref{full_ins}.

\section{Calibration}

PoGOLino is designed to autonomously perform accurate neutron count rate measurements in high photon and charged particle induced background environments. In this section, the relevant instrument performance tests will be described. The efficiency of the Phoswich-based background suppression system is described first. This is followed by a benchmarking test of the simulations using measurements of the absolute neutron counting rate. Lastly the autonomy and stability of the instrument design was tested using long duration measurements in varying temperature and pressure environments. 

\subsection{Phoswich system}

\begin{figure}[!b]
  \begin{center}
    \includegraphics[width=1\textwidth]{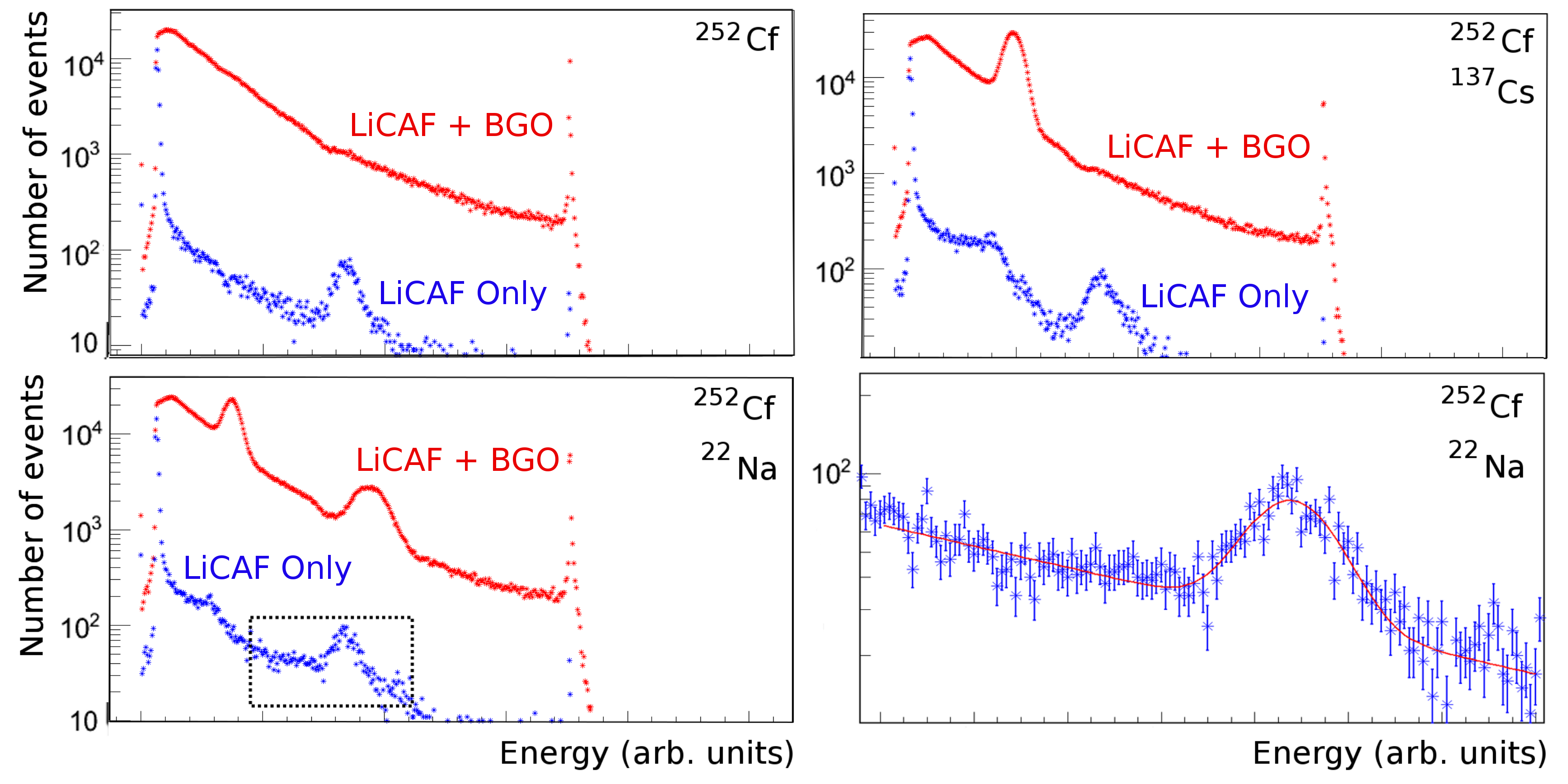}
  \end{center}
  \caption{The spectra as measured using the LiCAF+BGO PDC. All events from either BGO or LiCAF (in red) and only those recognised by the online veto system as coming from LiCAF (in blue) are shown for only $^{252}\mathrm{Cf}$ (top left), $^{252}\mathrm{Cf}$ with $^{137}\mathrm{Cs}$ (top right) and $^{252}\mathrm{Cf}$ with $^{22}\mathrm{Na}$ (bottom left). A zoom-in of the area within the black dotted line in the bottom left figure can be seen in the bottom right.}
  \label{Cfgamma}
\end{figure}

To test the ability of PoGOLino to accurately measure the neutron flux independently of the background using the online veto system, a set of calibration measurements was performed. The shielding of the PDCs by the passive materials of the instrument and the gondola will greatly reduce the background from low energy photons and charged particles, whereas minimum ionizing charged particles can be vetoed based on their large energy deposition. The main source of background is therefore expected to stem from photons with energies exceeding $100\,\mathrm{keV}$. An unshielded $^{252}\mathrm{Cf}$ source of which the neutron spectrum peaks around $1\,\mathrm{MeV}$ was placed at a distance of $16.5\,\mathrm{cm}$ from an unshielded LiCAF PDC. The $^{252}\mathrm{Cf}$ source used in the setup had an activity of 2.3 MBq corresponding to a neutron emission rate of $2.58\times10^5\,\mathrm{neutrons/s}$. Additionally $^{252}\mathrm{Cf}$ emits photons in the MeV energy range, the rate of which is a factor of $\sim2.7$ higher than the neutron emission rate. The total energy emitted in photons per fission reaction is $8.2\,\mathrm{MeV}$ which is distributed among the photons \cite{Cf_photon}, resulting in a broad spectrum in the MeV energy range. In order to further increase the photon flux and add emission lines to the photon spectrum a variety of $\gamma$ sources were added to the setup. Three measurements were performed, one without an additional $\gamma$ source, one with a $^{137}\mathrm{Cs}$ source, which has a strong emission line at $662\,\mathrm{keV}$, and one with a $^{22}\mathrm{Na}$ source, which has a strong emission line at $511\,\mathrm{keV}$ and a weaker one at $1274\,\mathrm{keV}$. The additional $\gamma$ sources were placed on top of the $^{252}\mathrm{Cf}$ source thereby maximizing the area of the LiCAF crystal unshielded by BGO. The photon background rate relative to the neutron signal rate during these measurements is one to two orders of magnitude higher than during flight. Figure \ref{Cfgamma} shows the spectrum from interactions in either BGO or LiCAF in red and the spectra as recognized by the online veto system as coming from the LiCAF alone in blue, together with an enlarged image of the fitted neutron capture peak from the $^{22}\mathrm{Na}$ measurement. The errors shown in the figure are the statistical errors of the number of counts per bin. The figure shows a significant increase in the rate when adding either of the two additional $\gamma$ sources to the setup. In the case of the additional $^{22}\mathrm{Na}$ source the neutron absorption peak can be observed to become less clear due to the $1274\,\mathrm{keV}$ emission from $^{22}\mathrm{Na}$, the absorption of which induces a pulse height similar to a neutron capture event. The dead time corrected neutron capture rate was acquired from the waveform data after application of the online veto with the use of a fit of the neutron absorption peak. The neutron count rate for irradiation with $^{252}\mathrm{Cf}$ only was found to be $4.38\pm0.26\,\mathrm{neutrons/s}$. For measurements with the $^{252}\mathrm{Cf}$ source in combination with $^{22}\mathrm{Na}$ and $^{137}\mathrm{Cs}$, the count rates were $3.98\pm0.21\,\mathrm{neutrons/s}$ and $4.79\pm0.29\,\mathrm{neutrons/s}$ respectively.
Despite the low signal-to-background ratio and the presence of strong emission lines in these measurements, the count rates as measured with additional gamma ray sources are within $1\sigma$ of the count rate measured with only $^{252}\mathrm{Cf}$. From these results it can be concluded that by using photon/neutron discrimination, the neutron flux can be measured accurately even in high radiation environments with a signal-to-background ratio one to two orders of magnitude lower than that expected for flight. 

\subsection{Absolute counting rate}

\begin{figure}[!b]
  \begin{center}
    \includegraphics[width=0.8\textwidth]{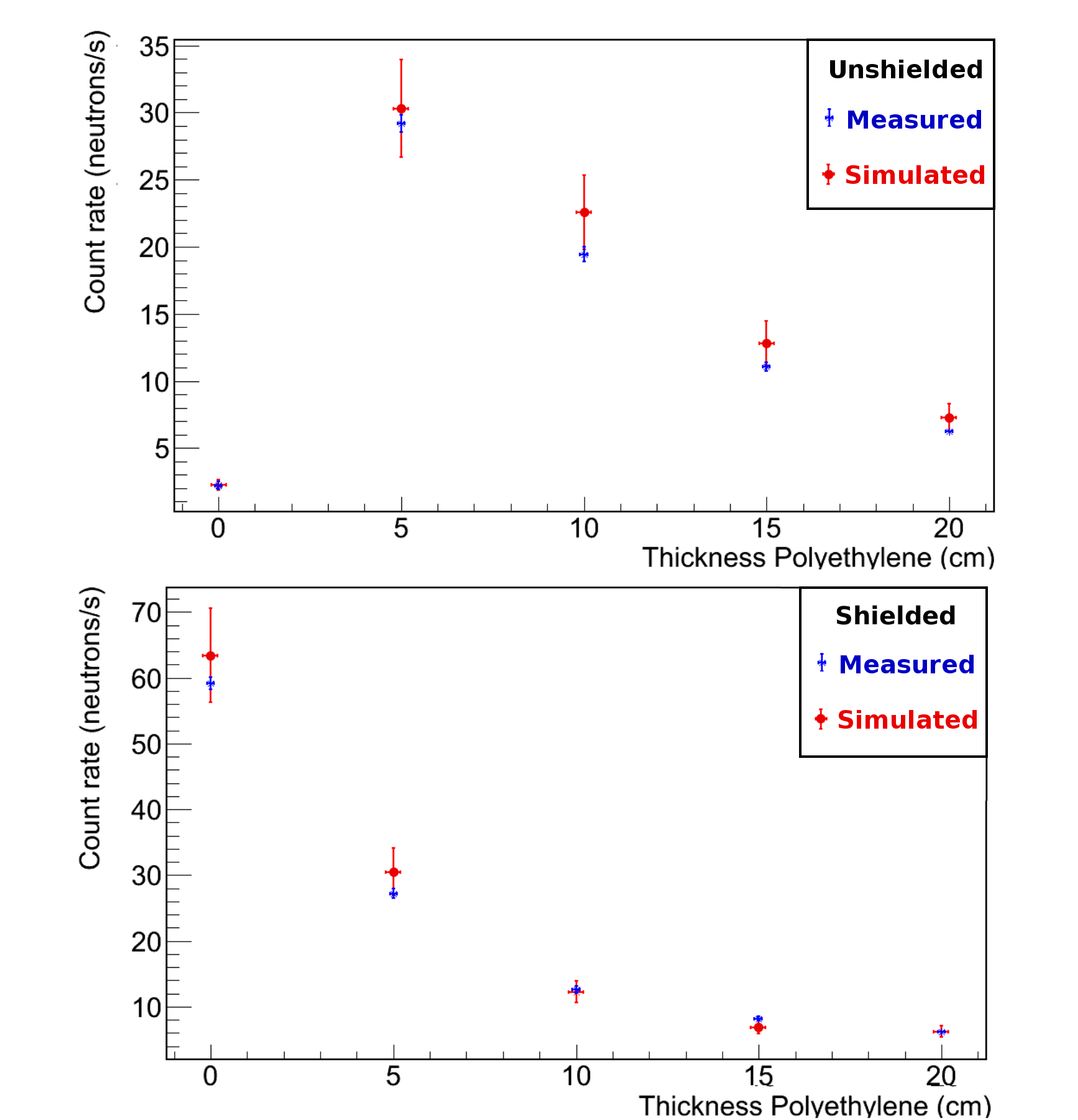}
  \end{center}
  \caption{The neutron count rate as measured (blue cross) and simulated (red dot) as a function of the thickness of the polyethylene moderator placed between the source and the detector both for the unshielded LiCAF PDC (top) and for the LiCAF PDC placed within an additional $7.5\,\mathrm{cm}$ cylindrical polyethylene shield (bottom).}
  \label{call_1}
\end{figure}

The two LiCAF PDCs were calibrated using simulations performed with Geant4.9.6. For the calibration, both the shielded and the unshielded detector were irradiated from a distance of 20 cm using a $^{252}\mathrm{Cf}$ source. The dead time corrected neutron capture rate was measured for both LiCAF PDCs with different amounts of additional polyethylene between the source and the detector. The measured count rate, taken as the integrated area under the neutron capture peak, was plotted against the thickness of the additional polyethylene. The results are presented in figure \ref{call_1} together with the counting rates simulated using Geant4, using the QGSP\_BERT\_HP physics list together with the high precision thermal neutron scattering data libraries, which are relevant for thermalisation in polyethylene. The measurement error is taken to be the error on the fit of the neutron absorption peak whilst the error on the simulated count rate is a systematic error, taken to be $10\%$, resulting from uncertainties in the measurement setup. To estimate this error, simulation parameters known with a poor precision, e.g. the thickness of the concrete roof, were varied within their uncertainty. The measured count rate as a function of polyethylene thickness can be seen to match the simulated event rates to within the $1\sigma$ errors, indicating that the neutron capture cross section by LiCAF and the neutron moderation by polyehtylene can be accurately simulated using Geant4.

\subsection{Instrument Performance Stability}

In order to test the instrument's capability to perform measurements autonomously over long periods of time with large variations in outside temperature and pressure, several tests were performed. Two parameters which are expected to vary with temperature during flight are the light yield and rise time of the BGO \cite{BGO_temp}. Depending on the atmospheric conditions, the temperatures encountered during the ascent from the Esrange Space Centre during March can range between $-10\,^\circ\mathrm{C}$ and $-80\,^\circ\mathrm{C}\,$ \cite{Kent}. Due to these extreme temperature changes during flight it is important to understand the temperature dependence of the waveform discrimination efficiency. The on-ground neutron background flux was therefore measured continuously in a climate chamber over a 3 day period, while the instrument was powered using an external power supply and the outside temperature was varied between room temperature and $-40\,^\circ\mathrm{C}$. Despite the possibility for temperatures below $-40\,^\circ\,\mathrm{C}$ at float altitude, the heat loss during these on-ground measurements is higher than during flight due to the presence of atmosphere during this test which causes additional heat loss through convection and conduction. Despite an observed increase in the decay time of BGO events, no difference in the neutron counting rate was found. The online waveform discrimination criteria can therefore be assumed to be valid down to outside temperatures of at least $-40\,^\circ\mathrm{C}$. Furthermore the electronics of the instrument were shown to operate as expected for a duration of 3 consecutive days without a need for outside intervention. The temperature inside the instrument was found to stabilise at a minimum of $0\,^\circ\,\mathrm{C}$ during the measurements, well above critical minimum temperatures of any of the used components. Lastly no significant change in the scintillation properties of LiCAF were observed.

To further test the temperature behaviour of PoGOLino the instrument was placed in a thermal vacuum facility at SSC in Solna, Sweden. The pressure was set to equal $5\,\mathrm{g/cm^2}$, the atmospheric pressure found at $\sim35\,\mathrm{km}$ altitude. The temperature in the climate chamber was set to the minimal possible temperature of $-20^\circ\,\mathrm{C}$. The temperature in the instrument stayed around $20^\circ\,\mathrm{C}$ and the pressure in the instrument was found to fluctuate with temperature as expected. A second test was performed during which the instrument was turned off in the thermal vacuum chamber in order for it to cool down, the instrument turned on successfully with a temperature of $-20\,^\circ\,\mathrm{C}$ inside. It can therefore be concluded that the instrument is pressure tight down to low temperatures and that the expected temperatures inside the instrument during flight will stay within the operational range in which no loss of background rejection efficiency is expected. 

\section{Flight}

\begin{figure}[!h]
  \begin{center}
    \includegraphics[width=0.5\textwidth]{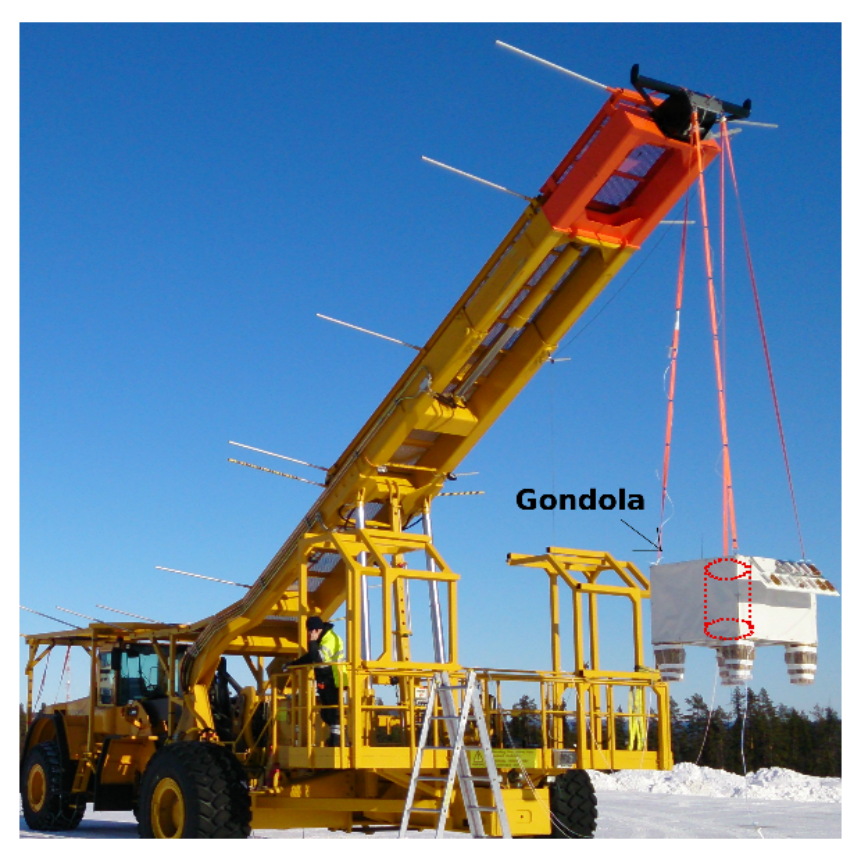}
  \end{center}
  \caption{The PoGOLino gondola before launch hanging from the launch vehicle. The position of PoGOLino is indicated with the dotted red lines.}
  \label{gondola}
\end{figure}

\begin{figure}[!h]
  \begin{center}
    \includegraphics[width=0.7\textwidth]{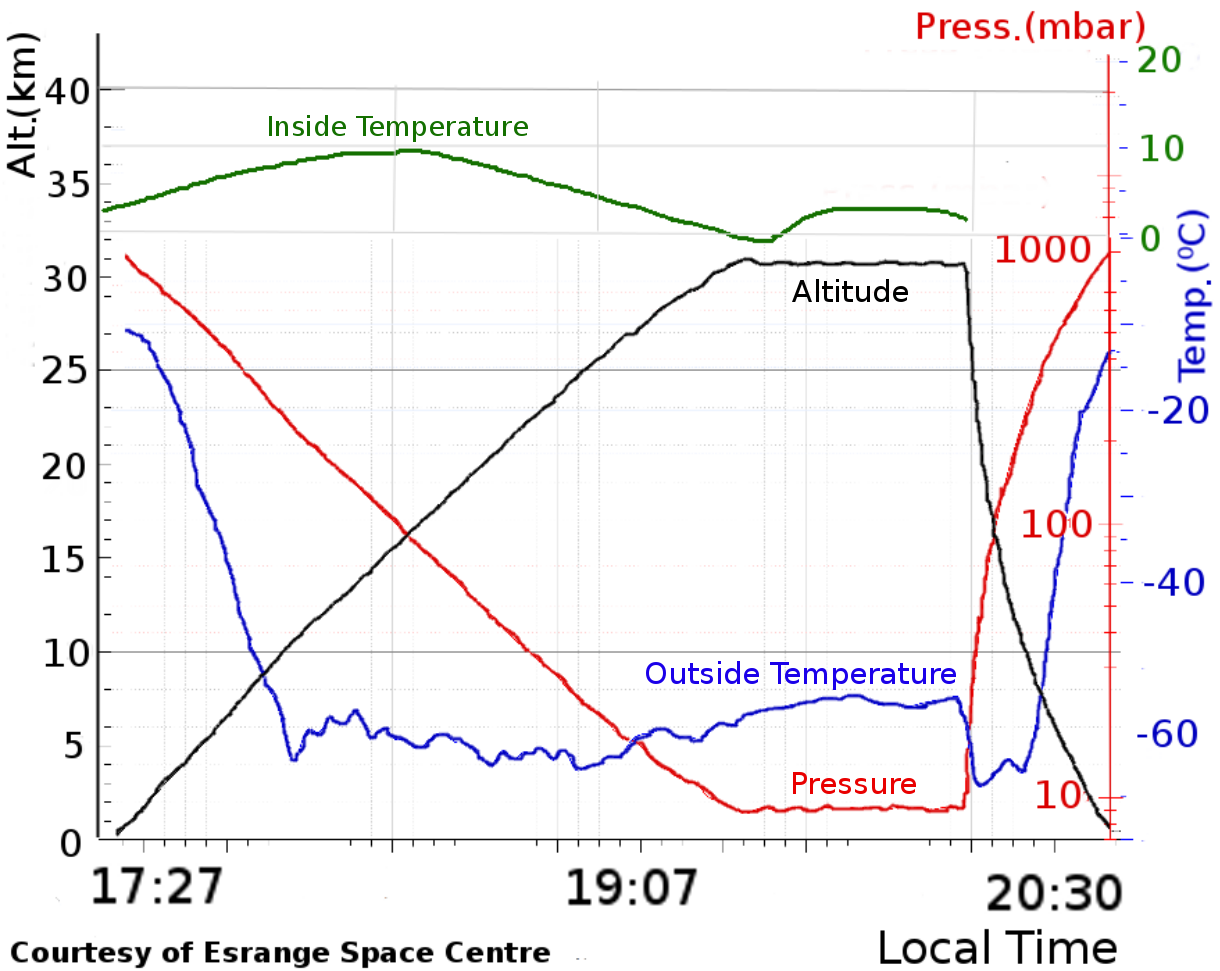}
  \end{center}
  \caption{The altitude (black), outside pressure (red) and outside temperature (blue) as measured in the gondola \cite{Esrange} and the temperature as measured inside the PoGOLino pressure vessel (green).}
  \label{flightdata}
\end{figure}

PoGOLino was launched at 17:27 LT on March 20th 2013 from the Esrange Space Centre in Northern Sweden, which is located at a magnetic latitude of $65.2^\circ$ degrees. The purpose of the main payload was flight qualification of different components by the Esrange Space Centre. PoGOLino was placed in the gondola as a piggyback experiment. The detector was positioned in a corner of the gondola as far away from all other payload equipment as possible, thereby reducing neutron moderation effects. The gondola, shown in figure \ref{gondola}, reached a float altitude of 30.9 km after an ascent phase of almost 2 hours. The subsequent time at  float was approximately one hour, after which the gondola was separated from the balloon using an explosive charge. After separation, the gondola started its parachute guided descent and returned safely to ground in Muonio, Finland, (magnetic latitude of $64.9^\circ$), about 100 km to the east of the launch site. The altitude, pressure and outside temperature are shown as a function of time in figure \ref{flightdata} together with the temperatures as measured inside the PoGOLino pressure vessel. The temperature inside the instrument can be seen to vary between $0\,^\circ\mathrm{C}$ and $10\,^\circ\mathrm{C}$ degrees, as expected.

\begin{figure}[!h]
  \begin{center}
    \includegraphics[width=1.0\textwidth]{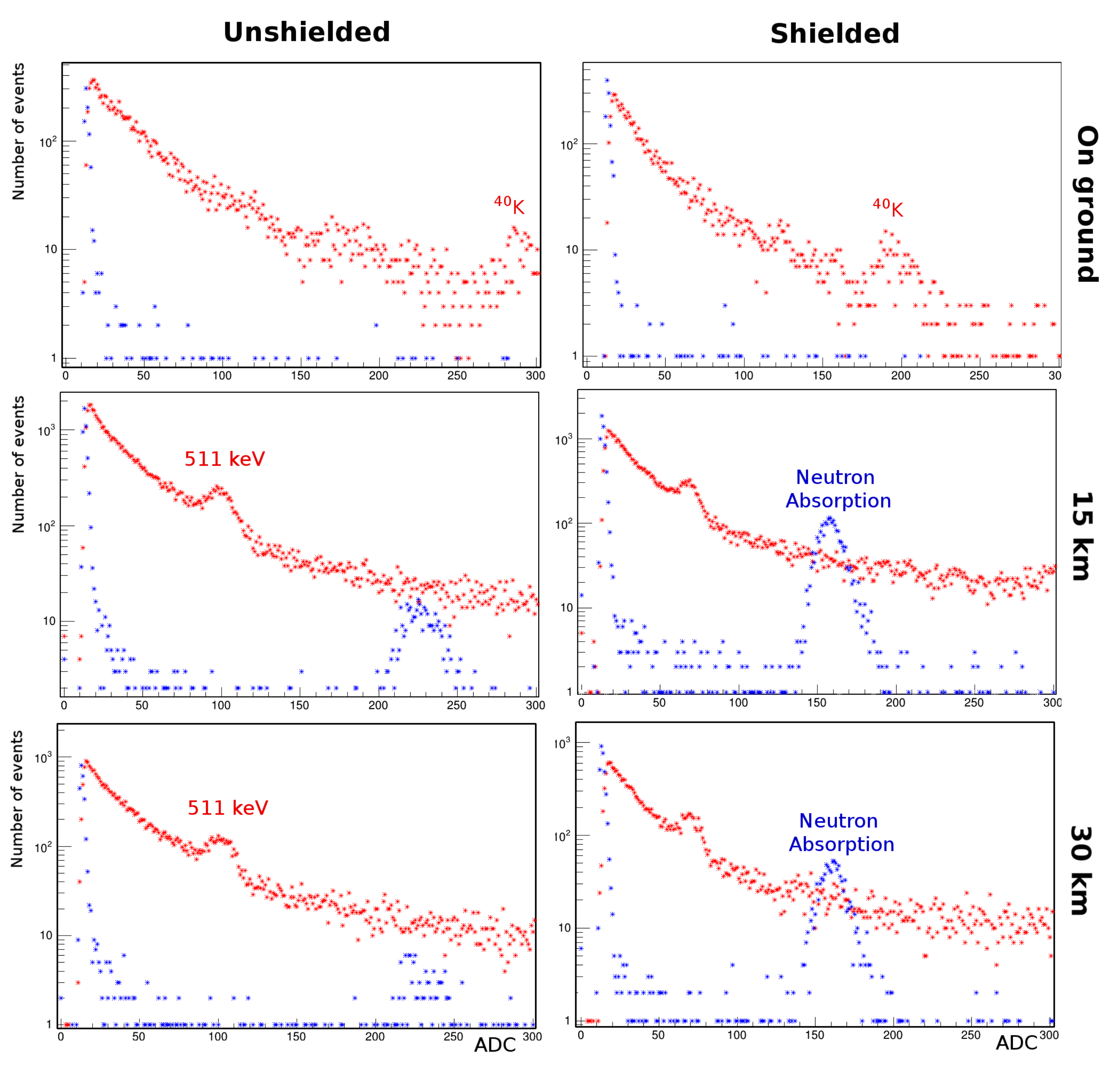}
  \end{center}
  \caption{Spectra as measured by the unshielded (left) and shielded PDC (right) for three different altitudes, on ground (top), 15 km (middle) and 30 km (bottom). The spectra, taken from the histogram data, as measured by BGO is shown in red and that from LiCAF in blue. On ground the BGO can be seen to measure emission for $\mathrm{^{40}K}$ decay, during flight this emission peak disappears, instead the 511 keV peak can be observed.}
  \label{flight_data}
\end{figure}

The LiCAF PDCs recorded data from half an hour before launch, in 5 minute runs, up to the moment the payload was cut from the balloon. Spectra from data taken on ground, at 15 km and at float by both PDCs are shown in figure \ref{flight_data}. The BGO data, shown in red, shows a clear peak caused by $511\,\mathrm{keV}$ photons in both PDCs in the flight spectra and a peak resulting from $\mathrm{^{40}K}$ in the spectra taken on ground. The LiCAF spectrum, in blue, shows the neutron capture peak. The photon induced background in the LiCAF is negligible around the peak. Two sets of data with vetoes off were performed towards the end of the flight when the altitude of the payload was stable. The counting rates as reconstructed from these data sets were in perfect agreement with the data taken with veto application.

\begin{figure}[!h]
  \begin{center}
    \includegraphics[width=1.0\textwidth]{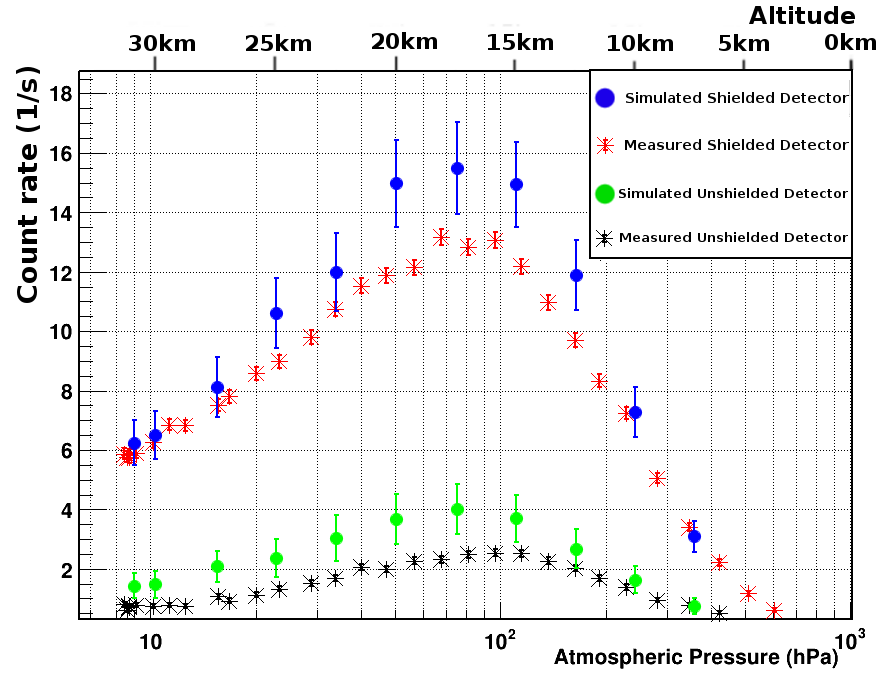}
  \end{center}
  \caption{The count rate as measured in the shielded (red cross) and unshielded (black cross) PDC compared to the count rates as simulated using spectra coming from \cite{Me!}, for the shielded detector (blue circles) and the unshielded detector (green circles) as a function of altitude (top axis) and atmospheric pressure (bottom axis).}
  \label{count_alt}
\end{figure}

The altitude dependence of the count rate as measured by both detectors is shown in figure \ref{count_alt} and compared to the count rate expected from simulations based on spectra taken from \cite{Me!} extrapolated down to thermal neutron energies. The upward and downward moving part of the neutron environment were simulated separately. The upward moving spectrum was emitted isotropically from the bottom half of an instrument centred sphere with a radius of $2\,\mathrm{m}$, while the downward moving spectrum was emitted isotropically from the top half of this sphere. For these simulations the instrument was placed in a detailed model of the flight gondola. Using results presented in \cite{Usoskin} and data from the Oulu Neutron Monitor \cite{Oulu} the solar activity during flight was estimated with a potential parameter of $\phi= 800\,\mathrm{MV}$ (the flight took place towards the end of a strong Forbush decrease). The count rate induced on the instrument by the neutron flux was simulated using Geant 4.9.6 using the physics lists validated using the ground calibration simulations. The temperatures were set equal to the temperatures measured during flight. The measured neutron count rates can be seen to be negligible at high pressures (low altitude) due to shielding effects of the atmosphere. When moving up in the atmosphere the counting rate increases down to pressures of around $50-100\,\mathrm{hPa}$ ($15-20\,\mathrm{km}$) where they reach a maximum at an altitude corresponding to the mean interaction altitude for charged cosmic rays. With increasing altitude the neutron flux decreases again towards the lowest measured pressures (below $10\,\mathrm{hPa}$). 

The measurements of the shielded detector can be seen to be in good agreement with the simulations in figure \ref{count_alt}. An overestimation on the order of $10\%$ by the simulations can be seen at the peak of the curve which is found at the altitude where most neutrons are produced. This discrepancy between the measured and simulated rate is most likely a result of the over-simplifications in the directional dependence of the simulated incoming flux. This results from treating the flux as comprising only an upward and downward moving part. The directly downward moving flux is therefore underestimated in the region where most neutrons are produced, whereas the horizontally moving flux is overestimated, resulting in an overall overestimation of the measured rate. At higher and lower altitudes there is no such discrepancy because of the higher isotropy of the directional dependence of the neutrons resulting from scattering interactions subsequent to their production. As a result the measured counting can be seen to agree well with the simulated rates for these altitudes.

For the unshielded detector the discrepancy between measurement and simulation is larger, this is most likely again a result from a simplification in the simulated spectra. This discrepancy is most likely a result of an oversimplification of the thermal part of the spectrum in the used spectral model, the model is based on simulation data for neutrons with energies above 1 keV only. The discrepancy can furthermore stem from uncertainties in the simulation of the payload gondola to which the unshielded detector is significantly more sensitive due to its high sensitivity in the thermal energy region. Both above and below the production region the measurements and simulations are found to be within $1\,\sigma$ of each other. Overall it can be concluded that the counts rates as measured by PoGOLino match the simulations well, these results furthermore show the need for more high altitude neutron data in order to improve currently existing atmospheric neutron models.

\section{Conclusion}

PoGOLino was designed to reduce the shortage of data on the neutron environment at high latitude balloon altitudes. Such data is essential for improving and validating atmospheric neutron models based on Monte Carlo techniques. The balloon-borne instrument makes use of LiCAF scintillator-based neutron detectors with BGO anticoincidence. PoGOLino was shown to accurately measure the neutron count rate over a wide energy range by making use of PDCs with and without polyethylene shielding. Ground calibrations have demonstrated the high efficiency of the instrument to perform photon/neutron discrimination. By comparing calibration data of the absolute counting rate with simulations it was shown that the instrument performance can be simulated accurately using Geant4. 

The flight of PoGOLino took place on March 20th 2013 during which it provided the first dedicated measurements of the neutron environment at high latitudes at stratospheric ballooning altitudes. The instrument measured the neutron environment up to an altitude of $30.9\,\mathrm{km}$ at a magnetic latitude of $65^\circ$. The measurement results show that by making use of the BGO-LiCAF PDC principle, the gamma-ray background on the neutron measurement can be reduced to a negligible level. The measured data was furthermore shown to be in good agreement with simulations. 

The use of BGO-LiCAF PDCs has provided PoGOLino with a high neutron capture efficiency and signal-to-background ratio both of which are needed to perform measurements in the highly variable, high radiation environment encountered at stratospheric altitudes. The instrument design furthermore ensures a low mass and complete autonomy, both of which make it suitable for small balloon flights. The PoGOLino results show the high potential for LiCAF-BGO PDCs to be used in future instrument designs. By increasing the number of PDCs and providing different shielding thickness’s to these channels such an instrument could be used to provide accurate neutron energy spectra without a dependency on $\mathrm{^3He}$.

\section{Acknowledgement}

The authors acknowledge: The Swedish National Space Board for funding the PoGOLino project; the SSC Esrange Space Centre for launch services and accepting PoGOLino on the TFB-01 test flight; the AlbaNova workshop for their work on the PoGOLino mechanics; Christian Lockowandt for organising the thermal vacuum test; the data from Oulu Neutron Monitor were provided by Sodankyl\"a Geophysical Observatory via \url{http://cosmicrays.oulu.fi/}.

\label{}








\end{document}